\documentclass[11pt]{article}
\usepackage[letterpaper, left=1in, top=1in, right=1in, bottom=1in,nohead,includefoot,ignoremp]{geometry}
\usepackage{charter,bm,graphicx,natbib,amsmath,amssymb,amsfonts,amsthm,hypernat}
\usepackage[usenames,dvipsnames,x11names]{xcolor}
\usepackage[center]{titlesec}
\usepackage{xy}\xyoption{all} \xyoption{poly} \xyoption{knot}  
 
\usepackage[pdftex,pagebackref=true]{hyperref}
\hypersetup{colorlinks,linkcolor=RoyalBlue1,urlcolor=RoyalBlue2, citecolor=RoyalBlue3}
\bibpunct{(}{)}{;}{a}{}{,}

\def\blu#1{{\color{RoyalBlue4}#1}}
\def\cD{\mathcal{D}}
\def\eqn#1{eqn.~(\ref{eq:#1})}

\begin{document}

\title{Bayesian Emulation for Optimization\\ in Multi-Step Portfolio Decisions}
\author{\blu{Kaoru Irie and Mike West}}
\maketitle

\begin{abstract}
We discuss the Bayesian emulation approach to computational solution of  
multi-step portfolio studies in financial time series.   {\em Bayesian emulation for
decisions} involves mapping the technical structure of a decision analysis problem to that of 
Bayesian inference in a purely synthetic  \lq\lq emulating" statistical model. This provides access to standard posterior analytic, simulation and optimization  methods that yield indirect solutions of the decision problem.
We develop this in time series portfolio analysis using classes of economically and psychologically relevant multi-step ahead portfolio utility functions. 
Studies with multivariate currency, commodity and stock index time series illustrate the approach and show some of the practical utility and benefits of the  Bayesian emulation methodology. 
\bigskip
\noindent {\em Some key words and phrases:} 
Bayesian forecasting; Dynamic dependency network models;  
Marginal and joint modes; Multi-step forecasting; Portfolio decisions; Synthetic model
\end{abstract}

\vfill \vbox{
\noindent\rule{3em}{.05em}\\
\footnotesize
 Kaoru Irie is Assistant Professor in the Department of Economics, University of Tokyo, Tokyo, Japan. \href{mailto:irie@e.u-tokyo.ac.jp}{irie@e.u-tokyo.ac.jp}.
Mike West is The Arts  \& Sciences Professor  of Statistics \& Decision
Sciences in the Department of Statistical Science, Duke University,
Durham, NC 27708. \href{mailto:mw@stat.duke.edu}{mw@stat.duke.edu}.

The research reported here was developed while Kaoru Irie was a graduate student 
in the Department of Statistical Science at Duke University. 
Kaoru was partly supported by a fellowship from the Nakajima Foundation of Japan, and he 
received the 2014-15  BEST Award for Student Research at Duke University. Support  from the Nakajima and BEST Foundations are gratefully acknowledged. 
Any opinions, findings and conclusions or recommendations expressed in this work are those of the authors and do not necessarily reflect the views of the Nakajima or BEST Foundations.

The authors thank Mengran Deng for sharing his unpublished undergraduate 
research findings (with M.W.)   related to the contents in Section \ref{sec:dlm}. 
}

\setcounter{page}{0}
\thispagestyle{empty}

\section{Introduction}
 
This work stems from an interest in Bayesian portfolio decision problems
with long-term, multi-step investment objectives that lead to the need for computational methods
for portfolio optimization.  Methodological advances reflect the fact that some such optimization problems can be recast-- purely technically-- as problems of computing modes of marginal posterior distributions in \lq\lq synthetic" statistical models.  We then 
have access to analytic and computational machinery for exploring posterior distributions 
whose marginal modes represent target optima in originating optimization/decision problems. We
refer to this as {\em Bayesian emulation for decisions}, with the synthetic statistical model
regarded as an emulating framework for computational reasons. 

The use of decision analysis for portfolios coupled with dynamic models for forecasting 
financial time series continues to be a very active area of Bayesian analysis-- in research and in
broad application in personal and corporate gambling on markets of all kinds.  Forecasting with 
multivariate dynamic linear/volatility models coupled with extensions of traditional Markowitz mean-variance optimization~\citep{markowitz} define benchmark 
approaches~(e.g. \citealp{pepe92,omar00,pepe03,pepe10, PolsonTew2000}, 
                         chapter 10 of \citealp{Prado2010}, \citealp{JacquierPolson2011}, among others). 
Much recent work has emphasised advances in forecasting ability based on increasingly structured
multivariate models~\citep[e.g.][]{Zhou2012,NakajimaWest2013,Nakajima2013jfe,Nakajima2014,Nakajima2016,ZhaoXieWest2015,GruberWest2015,GruberWest2016portfolios} with
benefits in portfolio outcomes based, in part, on improved characterizations of dynamics in multivariate 
stochastic volatility.  However, relatively little Bayesian work addresses interests in more relevant 
utility/loss functions, especially in longer-term, multi-step portfolio contexts; much of the cited work here
employs standard myopic/one-step ahead decision rules.  Our emphasis  is to complement 
these time series forecasting advances with Bayesian decision analysis that explicitly reflects 
personal or commercial utilities for stable portfolios in a multi-step context. 

In stylized forecasting and decision problems,    analysis involves
computing portfolio weight vectors to minimize   expected portfolio loss functions, and
to repeatedly apply this sequentially over time.  The solutions can be approximated numerically
in a number of ways, depending on the form of the loss function, but typically need 
customization of the numerical techniques.  The   approach here-- emerging naturally in the
specific context of multi-step portfolios--  is a general approach applicable to a variety of loss functions.
At any one time point with decision variable $w$ and 
expected loss function $L(w)$,  the Bayesian emulation strategy is useful if/when  
there exists a purely synthetic statistical model involving 
hypothetical random vectors (parameters, latent variables or states) $u,z$ and
generating a posterior density $p(u,z)$ under which the marginal model of $u$ is 
theoretically equal to the optimal $w$ in the portfolio decision. Minimizing $L(w)$ can then be 
approached by exploring   $p(u,z)$ with standard analytic and numerical methods
for posterior analysis.  While novel in terms of our context and  
development,  the basic idea here goes back (at least)  to~\citet{mueller1999}. There, with discrete decision variables in 
non-sequential design contexts,  optimization is solved using a similar synthetic posterior idea
and combining  optimization  with estimation using MCMC.  This approach has, surprisingly, seen limited
development, although recent work by~\cite{EkinPolsonSoyer2014} represents extension and new 
application.  Our work takes a broader emulating perspective with complete separation of models/forecasting  and decisions/optimization.    We develop  emulation of portfolio decisions 
using forecast information from state-of-the-art multivariate dependency network models~\citep{ZhaoXieWest2015}, treated as given. We then define the new multi-step decision  strategy for computing and revising Bayesian portfolios over time based on these forecasts. 

Section~\ref{sec:dlm} summarizes the multi-step portfolio set-up in sequential forecasting. To define and exemplify the emulation approach, we give summary details of 
its use in multi-step portfolios with extensions of standard (myopic, constrained) quadratic loss functions. 
Here the emulating synthetic statistical models 
are conditionally linear and normal state-space models, i.e., dynamic linear models (DLMs), 
amenable to evaluation using analytic forward filtering and backward smoothing (FFBS) methods. 
This is extended in  Section~\ref{sec:lasso} to a class of portfolios with sparsity-inducing 
penalties on portfolio weights and turnover.  The emulating models here also have 
state-space forms, but now with non-normal structure.   With 
augmented state-spaces, we can convert these to conditional DLMs in which posterior 
evaluation and mode search are efficiently performed by combining FFBS with a customized EM  method. 
Section~\ref{sec:marginal} discusses a fundamental question 
of definition of portfolio loss functions and objectives in multi-step contexts,
and a strategy for marginal mode 
evaluation. A range of portfolio loss functions are then 
evaluated in sequential forecasting and portfolio construction with a $13-$dimensional series 
of daily FX, commodity and 
market index prices. Section~\ref{sec:ex} discusses this,  
highlighting   choices of   portfolio loss functions and objectives, and  
practical benefits arising with sparsity-inducing, multi-step portfolio strategies. The latter shows 
the potential to improve portfolio outcomes, particularly 
in the presence of realistic transaction costs. Comments in Section~\ref{sec:conc} conclude the
main paper.    Appendices  provide technical details on 
optimization and on dynamic dependency network 
models used for forecasting. 

\textbf{Notational Remarks}: We use $p(x|y)$ for a generic   density of $x$ given $y.$ 
Normal, exponential and gamma distributions are written as $x\sim N(\mu ,\Sigma),$ 
$x\sim Ex(m)$ with mean $1/m,$ and $x\sim G(a,b)$ with shape $a$ and mean $a/b$; 
the values of their density functions  
at a particular $x$ are denoted by $N(x|\mu,\Sigma), Ex(x|m)$ and $G(x|a,b),$ respectively. 
Indices $s, s+1,\dots ,t$ for $s<t$ are shortened as $s{:}t$. 
The $k$-dimensional all-ones and all-zeros vectors are  $1_k = (1,\dots ,1)'$ and $0_k = (0,\dots ,0)'$, 
respectively, and $0$ represents a zero vector or matrix when dimensions are obvious.

\section{Multi-Step Emulation: Constrained Quadratic Losses \label{sec:dlm} } 
\subsection{Setting and Notation}
Over times $t{=}1,2,\ldots,$ we observe a 
$k-$vector asset price time series $p_t$; the returns vector $r_t$ has 
elements $r_{jt} = p_{jt}/p_{j(t-1)}-1,$  $(j=1{:}k).$  
At  time $t$ with current information set $\cD_t = \{ r_t, \cD_{t-1} \}$, 
a model defines a forecast distribution for returns at the next $h$ time points.   
With no loss of generality and to simplify notation, 
take current time $t{=}0$ with initial information
set $\cD_0.$  Predicting ahead, the predictive mean vectors and precision (inverse variance) 
matrices are denoted by 
$f_t = E[r_t|\cD_0]$ and $K_t = V[r_t|\cD_0]^{-1}$ over the $h-$steps ahead $t{=}1{:}h.$ 
The time $t$ portfolio weight vector  $w_t$ has
elements $w_{jt},$  some of which may be negative reflecting short-selling.
Standing at $t{=}0$  with a current, known portfolio $w_0,$  stylized myopic (one-step) Markowitz analyses are Bayesian
decision problems focused on choosing $w_1$ subject to constraints. Standard mean-variance 
portfolios minimize   $w_1'K_1^{-1}w_1$ 
subject to a chosen expected return {\em target} $m_1 = w_1'f_1,$ 
and usually a sum-to-one constraint $w_1'1_k=1$, i.e., allowing only portfolios     closed
to draw-down or additional investment. 

For multi-step portfolios, extend to consider the sequence of potential  
portfolio vectors $w_{1{:}h}$   over the next $h$ periods. 
The  decision is to choose $w_1,$ but we are  interested in target returns
and portfolio turnover control over multiple steps, and so must consider how the 
decision analysis might play-out up to time $t{=}h.$   Consider  multi-step (expected) 
loss functions of the form 
\begin{equation} \label{eq:loss1}
\begin{split}
 L(w_{1{:}h})&  \equiv  L(w_{1{:}h} |\cD_0) = \\
 \sum_{t=1}^h& \left\{ \alpha_t ^{-1}(m_t-f_t'w_t)^2
			     + \beta_t^{-1} w_t' K^{-1}_t w_t 
			     + \lambda_t^{-1}(w_t-w_{t-1})'W_t^-(w_t-w_{t-1})  \right\}
\end{split}
\end{equation}
where $\alpha_t$, $\beta_t$ and $\lambda_t$ are specified positive weights
defining relative contributions of the terms in this sum, while $W_t^-$ is the
(least-norm) generalized inverse of a specified $k\times k$
positive-semi-definite matrix  $W_t$, and will be the usual inverse in cases of 
positive-definiteness.    
Also,  $\cD_0$ now includes the current portfolio vector $w_0.$

The first set of terms in the sum involve specified multi-step target returns $m_{1{:}h}$.  
Individual investors typically prefer realized portfolios
to progress relatively smoothly towards an end-point target $m_h$, rather than bouncing from 
high to low interim returns. The weights $\alpha_t$ can be used to increasingly emphasize the
importance of later-stage returns as $t$ approaches $h.$ 
Note that allowing $\alpha_t\to 0$ theoretically implies the hard constraint on
expected return, $f_t'w_t = m_t$ as in the standard myopic case. 
Hence we refer to~\eqn{loss1} as including \lq\lq soft target constraints,'' while
having the ability to enforce the hard constraint at the terminal point via sending $\alpha_h$ to zero.
  
The second set of terms in~\eqn{loss1} penalize portfolio uncertainty using the standard risk measures
$V(w_t'r_t|\cD_0) = w_t'K^{-1}_tw_t$, again allowing differential weighting as a function of 
steps-ahead $t$.    The final set of terms relates to portfolio turnover.  If
$W_t=I_k$ these terms penalize changes in
allocations across all assets. If trades are at a fixed rate,   this is a direct transaction cost
penalty; otherwise, it still relates directly to transactions costs and so that terminology will be used.   
With a heavy emphasis on these terms-- as defined by the $\lambda_t$ weights--
optimal portfolios will be more stable over time,    providing less stress on investors
(including emotional as well as workload stress for individual investors). The   
  $W_t$ can play several constraint-related roles, as we discuss below.

\subsection{Portfolio Optimization and Emulating Models  \label{sec:dlmmodel} }

There are, of course, no new computational challenges to simple quadratic optimization implied by~\eqn{loss1}.
Key points are that it is easy to:  (i)  compute the joint optimizing values $w_{1{:}h}$, and 
(ii) deduce the  one-step optimizing $w_1$ for the Bayesian decision.  
Optimization with respect to $w_1$ alone
can be immediately performed using a forward-backward dynamic programming algorithm. 
Importantly, the optimizing value
for $w_1$ (or for any subset of the $w_t$) is-- as a result of the quadratic nature of~\eqn{loss1}--
precisely that sub-vector (or subset of vectors) arising at the global/joint maximizer $w_{1{:}h}.$ 
 
The emulation idea translates the above concepts into a synthetic Bayesian 
model immediately  interpretable by statisticians. Rewrite~\eqn{loss1} as
\begin{equation} \label{eq:dlmemulator}
e^{-\frac{1}{2}L(w_{1{:}h})} 
\ \propto \ \prod_{t=1}^h p(m_t| w_t ) p(z_t | w_t ) p(w_t | w_{t-1}) 
\equiv c \ 
 p(w_{1{:}h}|m_{1{:}h},z_{1{:}h},w_0)
\end{equation}
where each $p(\cdot|\cdot)$ term is a specific normal p.d.f., the $m_{1{:}h},z_{1{:}h},w_{1{:}h}$ are   interpreted as random quantities in a multivariate normal 
distribution underlying this density form, and 
where each $z_t$ is set at   $z_t=0.$    Specifically,  consider  a dynamic linear model (DLM) generating pairs of observations $(m_t,z_t)$-- with $m_t$ scalar and 
$z_t$ a $k-$vector--  based on latent $k-$vector states $w_t$ via
\begin{alignat}{3} 
  m_t &= f_t' w_t + \nu_t, &\qquad & \nu_t \sim N(0,\alpha_t), \label{eq:dlm1mt} \\
  z_t &= w_t + \epsilon_t, &\qquad & \epsilon_t \sim N(0,\beta_t K_t), \label{eq:dlm1zt} \\
  w_t &= w_{t-1} + \omega_t, &\qquad & \omega_t\sim N(0, \lambda_t W_t) \label{eq:dlm1wt}
\end{alignat}
with a known initial state (the current portfolio) $w_0$ and where the $\nu_t,\epsilon_s,\omega_r$ 
are independent and mutually independent innovations sequences. 
In this model, observing the sequence of synthetic observations $m_{1{:}h},z_{1{:}h}$ 
with $z_{1{:}h}=0$ immediately implies the resulting posterior  $p(w_{1{:}h}|m_{1{:}h},z_{1{:}h}=0,w_0)$
as given in~\eqn{dlmemulator}. 

Observe that computing the minimizer of $L(w_{1{:}h})$ is equivalent to 
calculating the posterior mode for $w_{1{:}h}$ in the synthetic DLM. It is immediate that the
required (marginal) optimizing value for $w_1$ is the marginal mode in this joint posterior. Since the
joint posterior is normal, marginal modes coincide with values at the joint mode, so we can 
regard the Bayesian optimization as solved either way.  We easily compute the
mode of $w_1$ using the forward filtering, backward smoothing (FFBS) algorithm-- akin 
to a Viterbi-style optimization algorithm~(\citealp[e.g.][]{Viterbi1967,Godsill2001,Godsill2004})-- widely
used in applications of   DLMs~\citep[e.g.][]{West1997,Prado2010}. 

\subsection{Imposing Linear Constraints \label{sec:constraints}}

As noted above, some applications may desire a hard target $m_h$ at the terminal point, and this
is formally achieved by setting the synthetic variance $\alpha_h=0$ in~\eqn{dlm1mt}. The 
multivariate normal posterior is singular due to the resulting constraint $m_h = f_h'w_h$, but this
raises no new issues as the FFBS computations apply directly. 
 
The general framework also applies with singular matrices $W_t$, now playing the roles of
the variance matrices of state innovations in~\eqn{dlm1wt}.  These arise to enforce linear portfolio constraints $Aw_t=a$ where $a$ is a given $n-$vector and $A$ is a full-rank  $n\times k$ matrix with $n<k.$ 
Choose $w_0$ to satisfy these constraints and ensure that each $W_t$ is such that $AW_t=0$.   Then 
the priors and posteriors for the synthetic states $w_t$ are singular and 
constrained such that $Aw_t=a$ (almost surely).  Again the FFBS analysis applies directly to generate 
the optimal portfolio vector $w_1$-- and the sequence of interim optimizing values $w_{1{:}h}$ even though only $w_1$ is used at $t{=}0$.  This now involves  propagating singular normal posteriors for states, 
as is standard in, for example, constrained seasonal DLMs~\citep[e.g.][sect.~8.4]{West1997}.
A key portfolio case is the sum-to-one constraint $1_k'w_t =1$ for all $t.$ 
Here we redefine  $W_t$ beginning with the identity $I_k$-- representing
equal and independent penalization of turnover across assets-- and then condition on the
constraints to give rank $k-1$ matrices $W_t \equiv W= I_k - 1_k1_k'/k.$

\section{Multi-Step Emulation: Constrained Laplace Losses \label{sec:lasso} } 

\subsection{Basic Setting \label{sec:laplacebasic}}
 
Now consider modifications  to  (i) more aggressively limit switching in/out of specific assets 
between time points-- for both transaction and psychological cost considerations, and 
to (ii) limit
the numbers of assets invested at any time point.  Several authors have considered 
absolute loss/penalties to encourage  shrinkage-to-zero of  optimizing portfolio 
vectors~\citep[e.g.][]{jagannathan2003risk,brodie2009sparse} and
we build on this prior work. Key points, however, are that such approaches have 
not been consistent with a Bayesian decision analysis framework, while 
goals with respect to marginal versus joint optimization in the multi-step context have 
been poorly understood and explored, and require clarification. 
Our fully Bayesian emulation strategy adds to this
literature while also clarifying this critical latter point and defining relevant methodology.

The \lq\lq Laplace loss" terminology relates to novel synthetic
statistical models that emulate portfolio optimization with absolute norm terms to penalize
portfolio weight changes. Modify~\eqn{loss1} to the form
\begin{equation} \label{eq:loss2}
\begin{split}
 L(w_{1{:}h})&  \equiv  L(w_{1{:}h} |\cD_0) = \\
 \sum_{t=1}^h& \left\{ \alpha_t ^{-1}(m_t-f_t'w_t)^2
			     + \beta_t^{-1} w_t' K^{-1}_t w_t 
			     + 2\lambda_t^{-1} 1_k' |w_t-w_{t-1}|\right\}
\end{split}
\end{equation}
where the final term now replaces the quadratic score with the sum of absolute changes 
of asset weights $1_k' |w_t-w_{t-1}| = \sum_{j=1{:}k}|w_{jt}-w_{j,t-1}|.$    
Relative to~\eqn{loss1}, this aims to more aggressively limit 
transaction costs, both monetary and psychological.  
Optimizing globally over $w_{1{:}h}$ may/will encounter boundary values in which 
some portfolio weights are unchanged between times $t-1$ and $t.$ This theoretical lasso-style
fact is one reason for the   interest in such loss functions, due to the implied 
expectation of reduced portfolio turnover-- or \lq\lq churn"-- and hence reduced costs.

\subsection{Emulating Dynamic Laplace Models \label{sec:lassomodel} }
 
In parallel to Section~\ref{sec:dlmmodel},   we   identify 
a synthetic statistical model-- again a state-space model but now with non-normal 
evolution/transition components for the synthetic latent states $w_t$-- of the form
\begin{alignat}{3} 
  m_t &= f_t' w_t + \nu_t, &\qquad & \nu_t \sim N(0,\alpha_t), \label{eq:ldlmm}\\
  z_t &= w_t + \epsilon_t, &\qquad & \epsilon_t \sim N(0,\beta_t K_t), \label{eq:ldlmz}\\
  w_{jt} &= w_{j,t-1} +\omega_{jt}, &\qquad & \omega_{jt}\sim L(\lambda_{t}^{-1}), & \qquad j=1{:}k, \label{eq:ldlmevo}
\end{alignat}
where $L(\lambda_{t}^{-1})$ denotes the Laplace (double exponential) distribution-- the
p.d.f. for each element is 
$p(w_{jt}|w_{j,t-1}) = \exp\{-|w_{jt}-w_{j,t-1}|/\lambda_t\}/(2\lambda_t)$. Also, 
the $\nu_t,\epsilon_s,\omega_{jr}$  are independent and mutually independent
across the ranges of all suffices. 

One of the immediate benefits of the Bayesian emulating model approach is that we can 
exploit latent variable constructs. In particular here,  the Laplace distributions are known to
be scale mixtures of normals~\citep{andrews1974scale,west1984a,west1987scale}.  Thus, there
exist latent random quantities $\tau_{jt}>0$, $(j{=}1{:}k, t{=}1{:}h),$ such that 
$\tau_{jt} \sim Ex(1/\{2\lambda_t^2\})$ independently over $j,t$, 
and based on which each synthetic state evolution in~\eqn{ldlmevo} 
has the form 
\begin{equation}
p(w_{jt}|w_{j,t-1}) = \int_0^\infty N(w_{jt}|w_{j,t-1},\tau_{jt})  Ex(\tau_{jt}|1/\{2\lambda_t^2\}) d\tau_{jt}.
\end{equation}  
Augmenting by the vectors of latent scales $\tau_t=\tau_{1{:}k,t}$, the  evolutions  
in~\eqn{ldlmevo} become
\begin{equation} \label{eq:dlm2}
  w_t = w_{t-1} + N(0, W_t), \quad W_t = \mathrm{diag}(\tau_t), \quad \tau_{jt} :iid\sim Ex(1/\{2\lambda_t^2\}). 
\end{equation}
This defines a conditionally normal DLM and the above/standard FFBS algorithm can be
used to evaluate the posterior mode of $p(w_{1{:}h}|W_{1{:}h},m_{1{:}h},z_{1{:}h})$ for any $z_{1{:}h}$ including that at zero. 
To maximize over portfolios $w_{1{}:h}$ in the implied marginal with respect to $W_{1:h}$, Bayesian EM~\citep[e.g.][]{dempster1977maximum}
is the obvious and easily implemented approach.  Here the E-step applies to the latent $W_{1{:}h}$,  while FFBS gives the exact
M-step for  $w_{1{:}h}$ at each iterate. In summary: 
\begin{enumerate} \setlength\itemsep{0pt}
 \item Initialization: Set each $w_t^{(0)}$ arbitrarily. 
 Candidates for initial values are   the  
current $w_0$, or  the trivially computed values that optimize the multi-step 
 portfolios under the quadratic loss of  Section~\ref{sec:dlm}. 
 
 \item For EM iterates $s=1{:}S$ under a chosen stopping rule, 
  repeat the following:.
 
 \begin{itemize}  \setlength\itemsep{0pt}
 
 \item {\em E-step:} For $j{=}1{:}k$ and $t{=}1{:}h,$ update $\tau_{jt}$   via  
$
\tau_{jt}^{(s+1)} = \lambda_t^2 | w_{jt}^{(s)}-w_{j,t-1}^{(s)} | 
$
to give a new matrix $W_t^{(s+1)}.$  

\item {\em M-step:} Implement FFBS for the emulating model of~\eqn{ldlmm} and (\ref{eq:ldlmz})
at $z_t=0$ and 
with augmented evolution in \eqn{dlm2}. This yields 
the exact mode $w_{1{:}h}^{(s+1)} $ of the synthetic  posterior conditional on
current $W_{1{:}h}^{(s+1)}.$
\end{itemize} 

\end{enumerate}
On stopping at iterate $S$, use $w_1^{(S)}$ as the approximate  
optimizing portfolio vector. 

The addition of linear constraints modifies the $W_t$ matrices  with details 
extending those of the normal model in Section~\ref{sec:constraints}.  
Write $V_t=\mathrm{diag}(\tau_t)$. Then for the full-rank set of $n<k$ constraints
$Aw_t=a$, the diagonal $W_t\equiv V_t$ is replaced  in~\eqn{dlm2} by  singular  
$W_t = V_t-V_tA'(AV_tA')^{-1}AV_t.$     In the key special case of sum-to-one constraints 
$1_k'w_t=1$ for all $t$, this reduces to  
$W_t = \mathrm{diag}(\tau_t) - \tau_t\tau_t'/(1_k'\tau_t).$

\subsection{Extended Laplace Loss for Sparser Portfolios \label{sec:nonneg} }

In the one-step, myopic context,  penalizing portfolio variance $w_1' K_1^{-1} w_1$ with a term 
proportional to  $1_k' |w_1| = \sum_{j=1{:}k}  |w_{jt}| $ is an obvious strategy towards the goal of 
inducing shrinkage to zero of optimized portfolio weights. As noted earlier, a number of recent works 
have  introduced such a lasso-style 
penalty directly on portfolio  weights, rather than on changes in weights, and with standard convex 
optimization algorithms for solution~\citep[e.g.][]{brodie2009sparse} and
demonstrating improved portfolio performance in some cases~\citep[e.g.][]{jagannathan2003risk}.
We now integrate such penalties as components of a more general class loss function embedded in the 
multi-step framework, and develop the Bayesian emulation methodology for this novel context. 

The shrinkage-inducing penalty  $1_k' |w_1|$ aims to drive some subset of weights to 
zero-- exactly in the one-step, myopic context when balanced only by portfolio risk. 
A key point to note is that,  when the portfolio vector is also subject to the sum-to-one
constraint, then the combined loss function also more aggressively penalizes negative 
weights, i.e., short positions, and so is particularly of interest to personal investors
and institutional  funds that generally adopt long positions.    That is, the absolute weight
penalty operates as a soft constraint towards non-negative weights. 
In our broader context below, this does not
theoretically imply non-negative optimal weights, but does often yield such solutions.
Modify~\eqn{loss2} to the form
\begin{equation} \label{eq:loss3}
\begin{split}
 L(w_{1{:}h})&  \equiv  L(w_{1{:}h} |\cD_0) = \\
 \sum_{t=1}^h& \left\{ \alpha_t ^{-1}(m_t-f_t'w_t)^2
			     + \beta_t^{-1} w_t' K^{-1}_t w_t 
			     + 2\gamma_t^{-1} 1_k' |w_t| 
			    + 2\lambda_t^{-1} 1_k' |w_t-w_{t-1}|\right\}
\end{split}
\end{equation}
with weights  $\gamma_t$ on the new absolute loss terms at each 
horizon $t=1{:}h.$  Extending the latent variable construction of double exponential 
distributions to these terms in addition to the turnover terms,  we now see that 
optimizing~\eqn{loss3}  is equivalent to computing the mode over states $w_{1{:}h}$ in 
a correspondingly extended synthetic DLM.  This emulating model is: 
\begin{align} \label{eq:dlm3}
\begin{split}
  m_t &= f_t' w_t + N(0,\alpha_t), \\
  z_t &= w_t + N(0,\beta_tK_t), \\
  u_t &= w_t + N(0,U_t), \quad \quad \ U_t = \mathrm{diag}(\phi_t), \quad \phi_{jt} :iid\sim Ex(1/\{2\gamma_t^2\}), \\
  w_t &= w_{t-1} + N(0, W_t),\quad 	W_t = \mathrm{diag}(\tau_t), \quad \tau_{jt} :iid\sim Ex(1/\{2\lambda_t^2\}), 
\end{split}
\end{align}
with synthetic  observations $m_t$ (scalar) and   $z_t{=}u_t{=}0$ ($k-$vectors), and
where latent scales $\tau_t$ are augmented with additional terms
$\phi_t = \phi_{1{:}k,t}$ for each $t.$ Conditioning on 
$\phi_{jt}$ converts the Laplace term $\exp(-|w_{jt}|/\gamma_t)$ to a conditional normal.
To incorporate exact linear constraints on each $w_t,$ the above is modified only through the implied
changes to the $W_t$; this is precisely as detailed at the end of Section~\ref{sec:lassomodel} above.
 
Extension of the FFBS/EM algorithm of Section~\ref{sec:lassomodel}
provides for computation of the optimizing $w_{1:h}$. 
Each E-step now applies to the latent $U_{1{}:h}$ as well as $W_{1{:}h}$,  
while the M-step applies as before to  $w_{1{:}h}$ at each iterate. 
Following initialization at $w_{1:h}^{(0)}$, the earlier details of 
iterates $s=1{:}S$ are modified as follows:

\begin{itemize}  \setlength\itemsep{0pt}
\item {\em E-step:}
\begin{itemize}  \setlength\itemsep{0pt}
\item Update the $\tau_{jt}$ via
$\tau_{jt}^{(s+1)} = \lambda_t^2 | w_{jt}^{(s)}-w_{j(t-1)}^{(s)} | $
to give a new matrix $W_t^{(s+1)}.$
\item 
Update the $\phi_{jt}$ via
$\phi_{jt}^{(s+1)} = \gamma_t^2  | w_{jt}^{(s)} | $ 
to give a new matrix  $U_t^{(s+1)}.$ 
\end{itemize} 
\item {\em M-step:}  FFBS applied to the extended emulating model~\eqn{dlm3} yields 
the exact mode $w_{1{:}h}^{(s+1)} $ of the synthetic  posterior conditional on
current $U_{1{:}h}^{(s+1)},  W_{1{:}h}^{(s+1)}.$
\end{itemize}  
The resulting  $w_1^{(S)}$ defines the 
optimizing portfolio vector.

\section{One-Step Decisions with Multi-Step Goals \label{sec:marginal}}
 
\subsection{Profiled Loss and Marginal Loss}
 
In  multi-step portfolio analysis, the  decision faced at time $t{=}0$ is to choose 
$w_1$ only. The future weights $w_{2{:}h}$ are involved in the initial specification of
the {\em joint} loss function  $L(w_{1{:}h})$  
in order to weigh expected fluctuations in risk and costs  up to the target horizon $t{=}h.$ 
From the viewpoint of Bayesian decision theory,  this is perfectly correct in the context of 
the actual decision faced if the approach is understood to be minimizing  
\begin{equation} \label{eq:profile}
 L(w_1) = \min\limits_{w_{2{:}h}} L(w_{1{:}h}). 
\end{equation}
Joint optimization over $w_{1{:}h}$ to deliver the actionable vector $w_1$ is 
Bayesian decision analysis with this implied loss as a function of $w_1$ alone.

The emulation framework provides an approach to computation, but also now suggests 
an alternative loss specification.  With emulating synthetic joint density $p(w_{1{:}h})$, 
minimizing the loss $L(w_1)$ above is equivalent to  {\em profiling out} 
the future hypothetical vectors $w_{2{:}h}$ by conditioning on their (joint) modal values. 
It is then natural 
to consider the alternative of {\em marginalization} over  $w_{2{:}h}$; that is, 
define the implied {\em marginal} loss function $L^{\ast}(w_1)$  as
\begin{equation} \label{eq:marginal}
 L^{\ast}(w_1) = -2 \log \left\{ p(w_1) \right\}, \qquad p(w_1) = \int p(w_{1{:}h}) dw_{2{:}h}.
\end{equation}
Call $L(w_1)$ the {\em profiled loss function} and $L^{\ast}(w_1)$ the {\em marginal loss function}.
 
In general, the resulting  optimal vectors $\hat{w}_1$ (profiled) and $w_1^\ast$ (marginal)
will differ.  A key exception is the case of the quadratic loss function and normal synthetic models of
 Section~\ref{sec:dlm} where the joint posterior $p(w_{1{:}h})$ is multivariate 
normal. In that case,  joint modes are joint means, whose elements are marginal means, i.e., 
$\hat{w}_1 = w_1^\ast.$   The situation is different in cases of non-normal 
emulating models, such based on the Laplace forms. These are now 
considered further for comparisons of marginal and profile approaches.

\subsection{Computing Marginal Portfolios under Laplace Loss \label{sec:margprof} }
 
Return to the Laplace loss framework of 
Sections~\ref{sec:laplacebasic} and \ref{sec:lassomodel} 
  (i.e., the extended Laplace context with $\gamma_t\to\infty$) 
with sum-to-one constraints.  Here the key issues of profiled versus marginal losses 
are nicely illustrated.  Similar features arise in the  
extended Laplace loss  context of Section~\ref{sec:nonneg}, but with no new conceptual or practical 
issues so details of that extension are left to the reader. 
The FFBS/EM algorithm easily computes the optimal profile portfolio ${\hat w}_1,$ but
it does not immediately extend to evaluating the optimal 
marginal portfolio $w_1^\ast$. Of several approaches   explored, the
most useful is based on   Markov Chain Monte Carlo (MCMC) analysis of the
synthetic DLM, coupled with iterative, gradient-based numerical
evaluation of the mode of the resulting Monte Carlo approximation to the required
marginal density function. Summary details are given here and further explored in
application in Section~\ref{sec:ex}.

The density $p(w_1)$ is the $w_1$ margin under the  full joint posterior of  
$( w_{1{:}h}, \tau_{1{:}h} )$ where $\tau_t = \tau_{1{:}k,t}$ is the vector 
of $t-$step ahead latent scales.  
The  FFBS/EM approach is enabled by the
nice analytic forms of implied conditional posteriors; these also 
enable MCMC analysis in this conditionally normal DLM with
uncertain scale factors. This approach is nowadays standard and easily implemented~(e.g. 
\citealp[][chapt.~15]{West1997}; \citealp[][sect.~4.5]{Prado2010}).   Now the FFBS is
exploited to generate {\em backward sampling}, rather than the backward smoothing that 
evaluates posterior modes. At each MCMC iterate, FFBS applies to {\em simulate} 
one draw of the full trajectory of states $w_{1{:}h}$  from the retrospective posterior 
$p(w_{1{:}h} |\tau_{1{:}h} )$ 
conditional on current values of  the latent scales. Then, conditional on this state trajectory, 
the conditional posterior 
$p(\tau_{1{:}h} | w_{1{:}h})$ 
is simulated to
draw a new sample of the latent scales. In the emulating model of~\eqn{dlm2} 
this second step involves a set of conditionally independent univariate draws, each from
a specific GIG (generalized inverse Gaussian) distribution. Applying the sum-to-one constraint on each $w_t$ vector changes this structure for the  $\tau_{jt},$ however, 
and direct sampling of the $\tau_{jt}$ is then not facile. To address this, we define
a Metropolis-Hastings extension for these elements to allow use of the constraint. 
Summary details of this, and of MCMC convergence diagnostics related to the 
real-data application in Section~\ref{sec:ex}, are given in Appendix~\ref{sec:appm}.

\newpage
The MCMC generate samples indexed by superscript $(i)$,  $i=1{:}I,$ for
some chosen sample size $I.$ The Rao-Blackwellized Monte Carlo approximation to 
the required margin for $w_1$ is then 
\begin{equation} \label{eq:mix}
\hat{p}(w_1) = I^{-1}\sum_{i=1{:}I} p( w_1  |\tau_{1{:}h}^{(i)}).
\end{equation}
Importantly,  this is the density of a mixture of $I$ normals: each 
conditional  $p( w_1 | \bullet)$ in the sum is the implied normal margin in the
DLM defined by conditioning values of latent scales, with moments trivially computable 
via FFBS (using backward smoothing), and the density values are easily evaluated at any $w_1.$ 
Thus the portfolio optimization problem reduces to mode-finding  in a mixture of multivariate normals,
and there are a number of numerical approaches to exploit for this.  The most effective is really 
one of the simplest-- a Newton-type updating rule based on the first order derivative of the density,
with repeat searches based on multiple initial values for numerical iterates. 
Relevant candidate initial values can be generated by evaluating the mixture at each of the normal component means, and selecting some of those with highest mixture density. Further details are noted 
in Appendix~\ref{sec:appm}.

\section{Studies in FX and Commodity Price Portfolios \label{sec:ex}}

\subsection{Data}
 Evaluation of multi-step portfolios uses data  on
daily returns of $k{=}13$ financial series: exchange rates of 10 international currencies (FX) 
relative to the US dollar, 
two commodities and two asset market indices; see Table~\ref{tab:list}. The time series runs 
from August 8, 2000 to December 30, 2011. An initial period of this data is used for exploratory analysis, followed by formal sequential filtering using a multivariate dynamic model, as noted below. The
main interest in portfolio evaluation is then explored over the period of $500$ days from 
January 1, 2009 to December 30, 2011.
\begin{table}[!htbp]
\begin{center}
\begin{tabular}{lclc}
Names & Symbol & Names & Symbol \\ 
\phantom{.}\\
Australian Dollar & AUD & Swiss Franc & CHF \\
Euro & EUR & British Pound & GBP \\
Japanese Yen & JPY & New Zealand Dollar & NZD \\
Canadian Dollar & CAD & Norwegian Kroner & NOK \\
South African Rand & ZAR & Oil price & OIL \\
Gold & GLD & Nasdaq index & NSD \\
S\&P index & S\&P & & \\    
\end{tabular} 
 \caption{\small Currencies, commodities and market indices.} \label{tab:list}
\end{center}
\end{table}%

\newpage

\subsection{Forecasting Model}
 
Forecasts are generated from a time-varying, vector auto-regressive model of order 
2~\citep[TV-VAR(2), e.g.][]{primiceri2005time,NakajimaWest2013}, 
with dynamic dependence network structure~\citep[DDN, ][]{ZhaoXieWest2015}.  
Exploratory analysis of the first 500 observations is used to define the sparsity structure of the
dynamic precision matrix for the TV-VAR innovations,  i.e., a sparse representation of
multivariate volatility, following examples in the above references.  From day 501, the analysis is 
run sequentially in time, updating and forecasting each day.  The DDN structure enables
analytic filtering and one-step forecasting; forecasting multiple steps ahead in a TV-VAR with
DDN structure is performed by direct simulation into the future.   For each day $t$ during the investment period,  
the model generates multiple-step ahead forecast mean vectors and variance matrices, 
$\{ f_{t+i},K^{-1}_{t+i} \}_{i=1{:}h}$,  given as Monte Carlo averages of $50{,}000$ forecast 
trajectories of the return vectors $r_{t+(1{:}h)}$ simulated  at time $t.$   We take $h{=}5$ days as the 
portfolio horizon, and reset the time index so that $t{=0}$ represents the start of the 
investment period, January 1, 2009.
Appendix~\ref{sec:ddn} provides detailed discussion of the DDN model, use of exploratory training data,  filtering and  simulation-based forecasting.

\subsection{Parameters and Metrics}
 
Comparisons use various values of portfolio parameters in 
 the quadratic/normal and  Laplace loss 
frameworks.  
In all cases, we take the target return schedule $m_{1{:}h}$ 
to be constant, with $m_t = 0.0005$ representing daily return targets of $0.05\%$, annualized
(261 trading days) to  about  $13.9\%.$       
Then, we have $\alpha_t>0$ for $t<h,$ rather than strictly enforcing the constraint by $\alpha _t = 0,$ 
so that the interim targets are \lq\lq soft" rather than strictly enforced. 
The initial portfolio $w_0$ is the myopic Markowitz portfolio for comparison. 
Parameters 
$\alpha_t, \beta_t, \lambda_t, \gamma_t$ define relative weights of 
the four components of loss. In a long-term context (e.g., when $t$ indexes months or more)
some use of discounting into the future becomes relevant. For example, 
we may take $\beta_t, \lambda_t, \gamma_t$
may be chosen to increase with $t$, but  $\alpha_t$ to decrease with $t$ to more aggressively 
target the soft targets as $t$ approaches $h,$ given the accurate and reliable long-term predictions.  
In short-term contexts,  such as with our $5-$day context, this
is not relevant, so we take constant weights 
$\alpha_t=\alpha=1, \beta_t=\beta, \lambda_t=\lambda, \gamma_t=\gamma.$ Setting $\alpha=1$ loses no loss of generality, as the remaining three weights are relative to $\alpha.$ Examples use 
various values of  $\beta, \lambda, \gamma$ to highlight their 
impact on portfolio outcomes. Larger values of $\beta$ reduce the penalty for risk in terms
of overall portfolio variance; larger values of  $\lambda_t$ leads to more volatile portfolio dynamics 
due to reduced penalties on transaction costs; and larger values of $\gamma$ reduce  shrinkage of
portfolio weights, also relaxing the penalties on shorting. 

Portfolios are compared in several ways, including realized returns. 
With a fixed transaction cost of $\delta\ge 0,$  time $t$ optimal portfolio vector  $w_t$ and realized return vector $r_t,$  cumulative return $R_t$ from the period $0{:}t$ is 
$  
 R_t= -1+\prod_{s=1:t}\{  (r_t+1_k)'w_t-\delta 1_k'|w_t-w_{t-1}| \}.
$  
In our examples, we compare cases with $\delta=0$ and $\delta=0.001.$ 
We also compare our multi-step portfolios with the standard one-step/myopic Markowitz approach--
naturally expected to yield  higher cumulative returns with no transaction costs as it then generates 
much more volatile changes in portfolio weights day-to-day.  Our portfolios constraining turnover 
are naturally expected to improve this in terms of both stability of portfolios and 
cumulative return in the presence of practically relevant, non-zero $\delta$.   Additional metrics
of interest are portfolio \lq\lq risk'' as traditionally measured by the realized portfolio
standard deviations $(w_t'K_t^{-1}w_t)^{1/2}$, and patterns of volatility in trajectories of
optimized portfolio weights over time.

\subsection{Normal Loss Portfolios \label{sec:exnormal} }

\begin{figure}[!htbp]
\centering
    \includegraphics[width=3.7in]{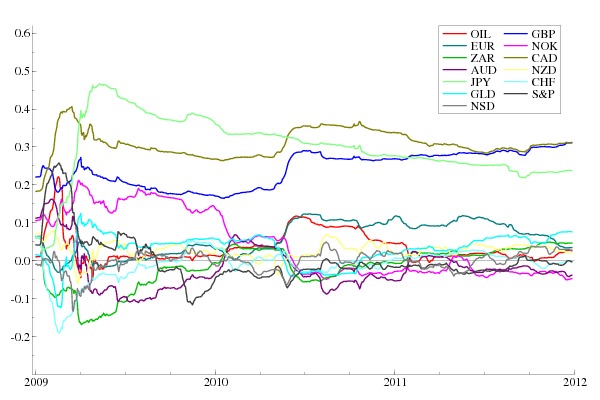}
    \includegraphics[width=3.7in]{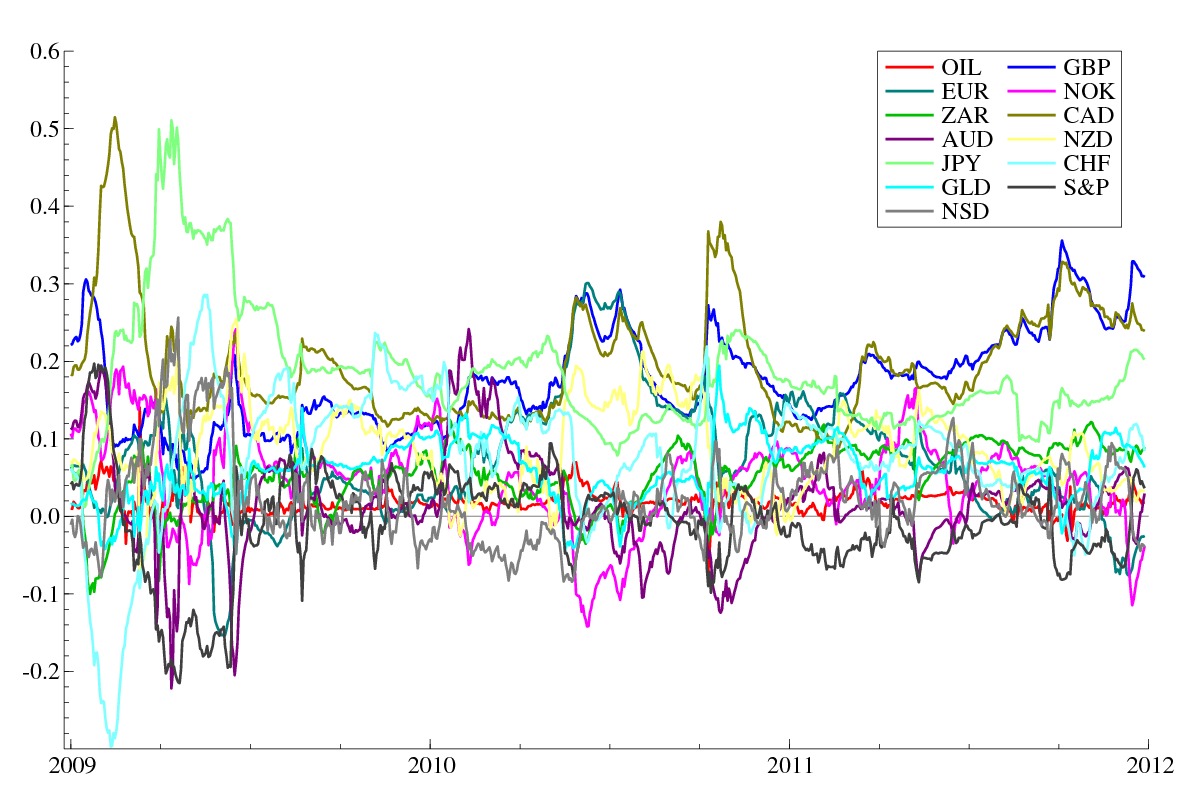}
    \includegraphics[width=3.7in]{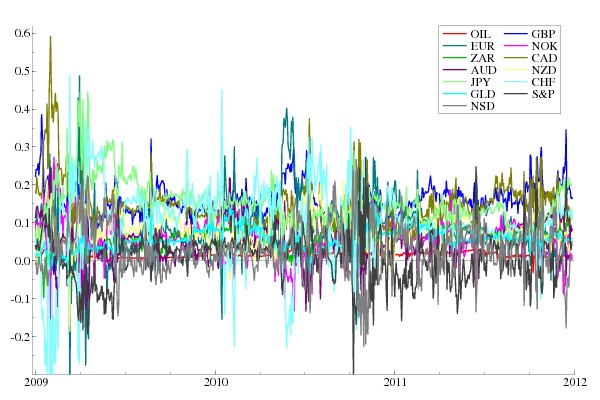} 
 \caption{\small Trajectories of optimal portfolio weights under normal loss 
with $\beta=100$,  $\lambda= 100$
(upper) and $\lambda=10{,}000$ (center), compared to traditional Markowitz weights
  (lower).} \label{fig:dlm} 
\end{figure}

First examples use the  normal loss framework of Section~\ref{sec:dlm} with  $\beta=100.$ 
Figure~\ref{fig:dlm} shows trajectories over time of optimized portfolio weight vectors using
$\lambda=100$ and $\lambda=10{,}000,$ as well as those from the standard, 
myopic Markowitz analysis that corresponds to $\lambda\to\infty.$  We see 
the increased smoothness of changes as $\lambda$ decreases; at $\lambda=1$ 
the trajectories (not shown) are almost constant.

\begin{figure}[!htbp]
\centering
    \includegraphics[width=3.7in]{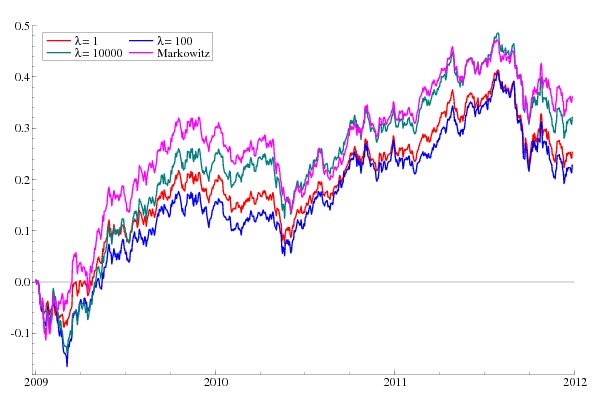}
    \includegraphics[width=3.7in]{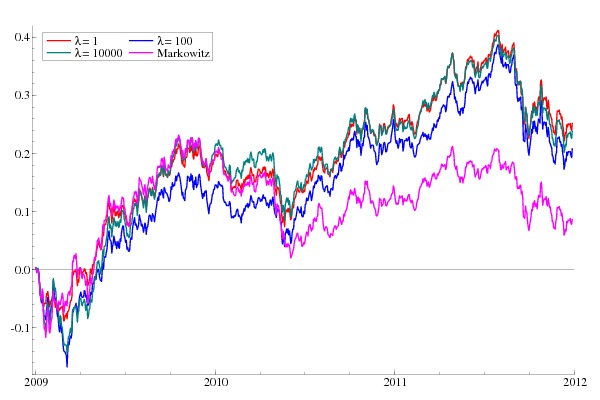} 
 \caption{\small Cumulative returns from normal loss portfolios with  $\beta=100$
and $\lambda= 1$ (red), 100 (blue), 
10{,}000 (green), together with Markowitz portfolios (pink). The
 transaction cost is $\delta=0$ (upper) and 0.001 (lower).  } \label{fig:dlmr}
\end{figure}

Figure~\ref{fig:dlmr} plots trajectories of cumulative  returns for three normal loss portfolios 
$(\lambda=1, 100$ and 10{,}000) and for the Markowitz analysis. 
Markowitz and larger $\lambda$ normal loss portfolios performs best-- in this metric-- with no transaction costs;  but the Markowitz approach is 
very substantially out-performed by the smoother, multi-step portfolios under 
even very modest transaction costs $(\delta=0.001).$  
Smaller $\lambda$ induces portfolios more robust to transaction costs. 
Of note here is that, during 2009 following the financial crisis, portfolios with 
larger $\lambda$ benefit as they are less constrained in adapting; but, later into 2010 and 2011, portfolios with 
lower $\lambda$ are more profitable as they define ideal allocations with less switching and therefore 
save on transaction costs. 

\begin{figure}[!htbp]
\centering
    \includegraphics[width=3.7in]{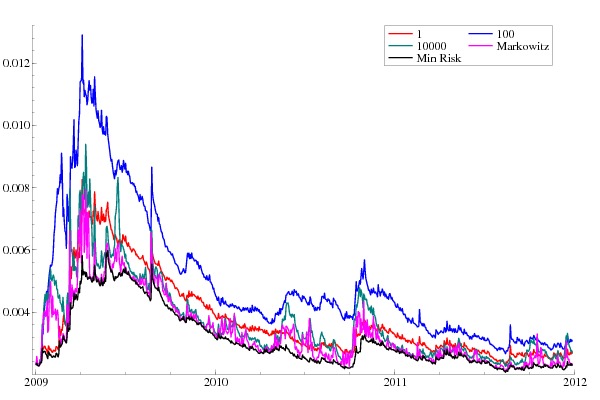}
 \caption{\small Trajectories of optimal portfolio standard deviations using normal loss with $\beta=100$, 
$\lambda= 1$ (red), 100 (blue), 
10{,}000 (green), the Markowitz portfolio (pink), and the minimum variance portfolio (black). } \label{fig:sd}
\end{figure}

Figure~\ref{fig:sd} shows trajectories of realized SDs of optimized portfolios, i.e. 
$(w_t'K_t^{-1}w_t)^{1/2}$ over time, for each of the portfolios in Figure~\ref{fig:dlmr}; 
also plotted is the theoretical lower bound 
trajectory $( 1_k'K_t^{-1}1_k )^{-1/2}$  from the  myopic, one-step, minimum variance portfolio. 
Less constrained portfolios with larger $\lambda$ have lower standard deviations, approaching those
of the Markowitz portfolio while also generating smoother changes in portfolio weights and higher cumulative 
returns. Thus, these portfolios are improved in these latter metrics at the cost of only modest 
increases in traditional portfolio \lq\lq risk." Interestingly, the relationship between $\lambda$ and 
realized standard deviations is not monotone;  
we see larger standard deviations in the case of $\lambda= 100$ than with $\lambda =1$, 
the latter, very low value yielding an almost constant portfolio over time that, in this study, turns
out to control risk at a level not matched by modestly more adaptive portfolios. 

\begin{figure}[!htbp]
\centering
    \includegraphics[width=3.7in]{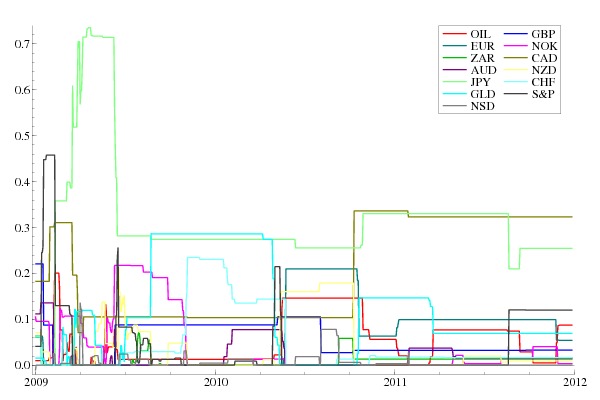}
    \includegraphics[width=3.7in]{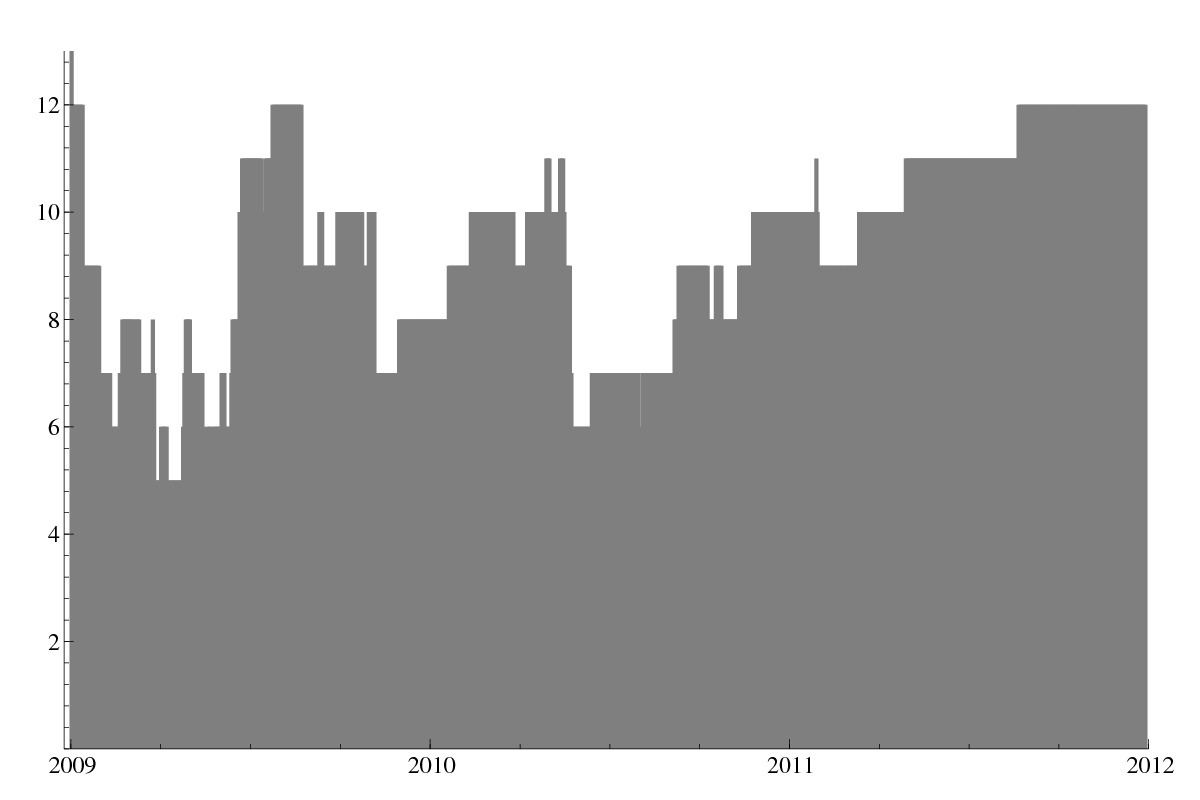} 
 \caption{\small  Trajectories of optimal portfolio weights (upper) and 
number of non-zero portfolio weights (lower) using 
extended Laplace loss  with $\beta=100$,  $\lambda= 100$ and $\gamma=100$.} 
\label{fig:em2}
\end{figure}%

\subsection{Extended Laplace Loss Portfolios}

We explore similar graphical summaries from analyses using the extended Laplace loss framework of 
Section~\ref{sec:nonneg}, and with sum-to-one constraints.
Figure~\ref{fig:em2} shows   optimal weight trajectories  
with $\alpha = 100$ and $\beta =1$ (this change of $\alpha$ is only in this example), and 
with the same level of penalization of turnover and absolute weights, i.e., 
$\lambda = \gamma=100.$ 
We see expected effects of the two types of shrinkage-- of changes in weights and in 
weights themselves. 
First, the hard shrinkage of changes induces much less switching in portfolio allocation over time, 
with longish periods of constants weights on subsets of equities.  This occurs even with 
larger $\lambda$ where the portfolio becomes more volatile and similar to  the Markowitz case. 
Second, the penalty on absolute weights themselves, and implicitly on short positions as a result in this
context of sum-to-one weights, yields trajectories that are basically non-negative on all equities over time. 
The joint optimization drives some of the weights exactly to zero at some periods of time, 
indicating a less than full portfolio over these periods.  Furthermore,  it is evident that there are 
periods where some of the weights-- while not zero-- are shrunk to very small values, so that
a practicable strategy of imposing a small threshold would yield sparser portfolios-- i.e., a \lq\lq soft" 
sparsity feature. 
Values $\lambda > \gamma$ favor more stability/persistence in the 
portfolio allocations, and we see more \lq\lq stepwise" allocation switches rather than more volatile
turnover. 
Conversely, $\lambda \le \gamma$ more aggressively favors no-shorting and encourages \lq\lq soft" 
sparsity  of allocations, resulting in dynamically switching portfolio weights over, generally, fewer assets.

\newpage

\begin{figure}[!htbp]
\centering
    \includegraphics[width=3.7in]{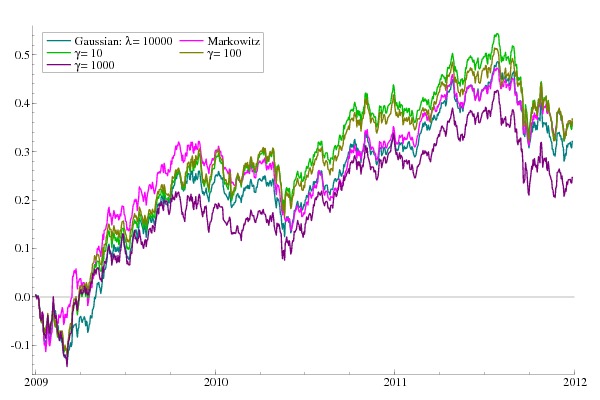}
    \includegraphics[width=3.7in]{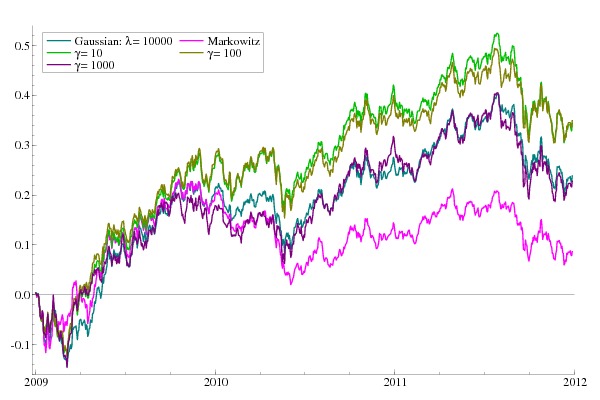} \\
 \caption{\small  Cumulative returns from extended Laplace loss portfolios with
$\beta=100,$  $\lambda= 100$ and $\gamma=$10 (green), 100 (yellow) and 10{,}000 (purple),
together with those from the normal loss portfolio with $\lambda=10{,}000$ (blue) and the Markowitz outcomes (pink).  
 The transaction cost is $\delta=0$ (upper) and 0.001 (lower). 
} \label{fig:em2r}
\end{figure}

Figure~\ref{fig:em2r} plots trajectories of cumulative  returns for three extended Laplace 
loss portfolios to show variation with the value of $\gamma,$ together with one highly adaptive normal loss portfolio and the Markowitz analysis, 
for $\alpha = 1$ and $\beta =100$ fixed. Again we compare cases with transaction cost  
$\delta=0$ and 0.001.  As with normal loss comparisons, all multi-step cases dominate the
traditional Markowitz analysis under even modest transaction costs.  In addition, we now see the
ability of the increasingly constrained multi-step Laplace portfolios to outperform unconstrained-- and hence
more volatile-- Markowitz as well as multi-step  normal loss portfolios even when transactions costs are zero
or ignored.  Then,  cumulative returns with $(\lambda,\gamma) = (100,1{,}000)$ are essentially uniformly dominated by those with $(\lambda,\gamma) = (100,10)$ and $(100,100)$, regardless of the existence of transaction costs in this example.  This suggests values of 
 $\gamma$ smaller than or comparable to $\lambda$ to appropriately balance the two
degrees of shrinkage while maintaining relevant returns. 
One  reason for this is the encouragement towards less volatile swings in weights
to larger negative/positive values and towards no-shorting as part of that, features that 
can lead to increased risk and transaction costs.

\subsection{Comparison of Profiled and Marginal Loss Approaches}

\begin{figure}[!htbp]
\centering
    \includegraphics[width=3.7in]{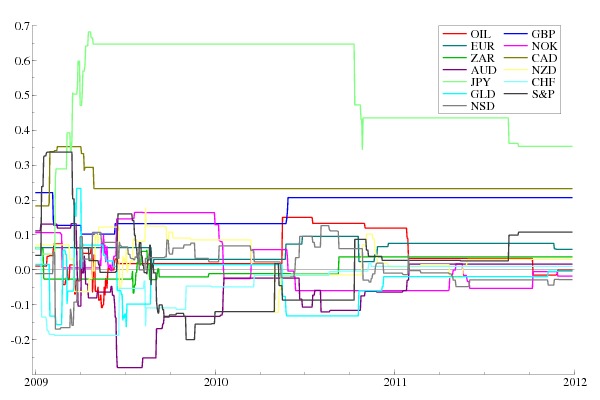}
    \includegraphics[width=3.7in]{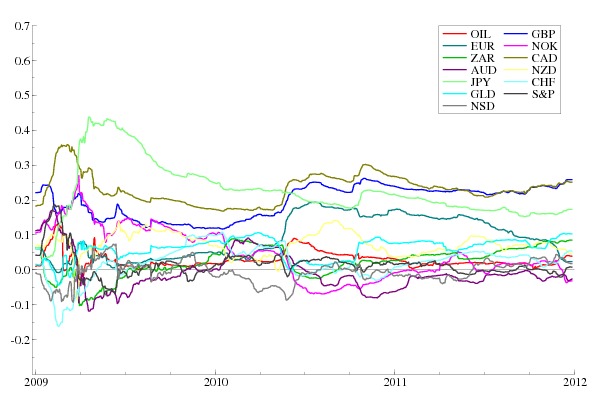} 
 \caption{\small Trajectories of optimal portfolio weights under 
Laplace loss  with $\beta=100$ and  $\lambda= 100$, showing 
profiled weights (upper) and marginal weights (lower).  
} \label{fig:ma}
\end{figure}

We now discuss some analysis summaries related to the discussion of profiled and marginal 
losses of Section~\ref{sec:marginal}.  As discussed in Section~\ref{sec:margprof} we do this in the 
Laplace loss framework of  Sections~\ref{sec:laplacebasic} and \ref{sec:lassomodel}
 (i.e., with $\gamma_t\to\infty$ in the extended context).  First,  Figure~\ref{fig:ma} shows 
 optimal weight trajectories with $\beta=100$ 
and $\lambda=100,$  comparing  the  profiled Laplace loss weights ${\hat w}_\bullet$ of
Section~\ref{sec:lassomodel} with the marginal Laplace loss weights $w_{\bullet}^\ast$ of 
Section~\ref{sec:marginal}.  Both strategies generate positive
weights on the JPY, GBP and CAD FX rates, with a number of the other assets having 
quite small weights for longer periods of time, while the weights under profiled loss vary more widely to 
higher absolute values. We see constant weights for long periods on 
adaptively updated  subsets of assets using the profiled weights, as expected; these trajectories are 
effectively smoothed-out and shrunk towards zero under the marginal weights.  The latter do 
not  exhibit the exact zero values that the former can, as we now understand is 
theoretically implied by our representation via the emulating statistical model: marginal modes
will not be exactly at zero even when joint modes have subsets at zero. 
 
\begin{figure}[!htbp]
\centering
    \includegraphics[width=3.2in]{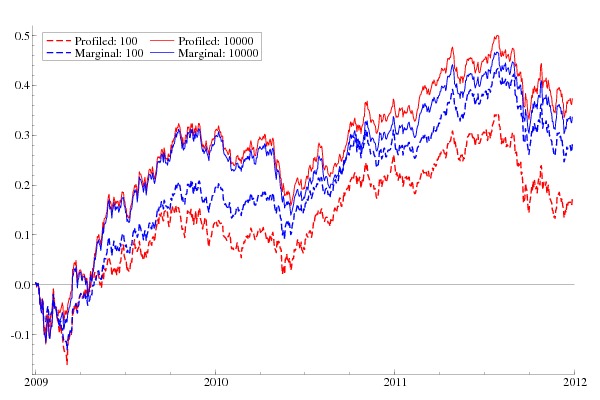}
    \includegraphics[width=3.2in]{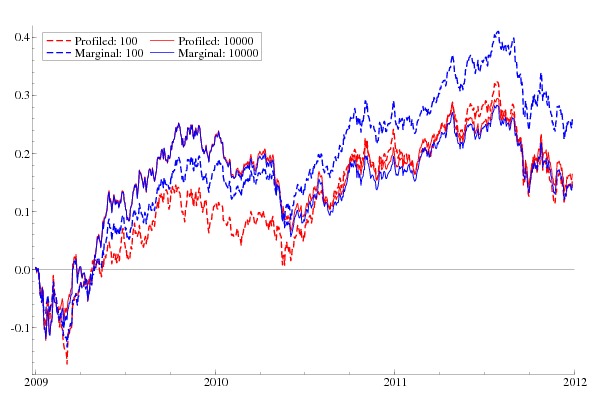} \\
 \caption{\small  Cumulative returns with extended Laplace loss,  comparing 
 [profiled, $\lambda=100]$ (red, dashed), 
 [profiled, $\lambda=10{,}000]$ (red, full),  
 [marginal, $\lambda=100]$ (blue, dashed), and  
 [marginal, $\lambda=10{,}000]$ (blue, full). 
 The transaction cost is $\delta=0$ (upper) and 0.001 (lower). 
 }
\label{fig:mar}
\end{figure}

Figure~\ref{fig:mar} plots trajectories of cumulative  returns for both profiled and
marginal portfolios in each of the cases with $\lambda=100$ and $\lambda=10{,}000.$ 
Without transaction cost, the profiled and marginal portfolios are similarly lucrative 
whatever the value of $\lambda$,  whereas the profiled portfolios show greater differences. 
In contrast, both approaches are more substantially impacted by transaction costs and
in a similar way; the cumulative return performance of portfolios  
decreases  drastically in the presence of transaction costs. 
Not shown here,  portfolios with smaller $\lambda$ values define far more 
stable weights while resulting in very similar cumulative returns under both
profiled and marginal strategies, as the resulting portfolio weights are very stable over time; this
extends this observation as already noted in the  normal loss context in Section~\ref{sec:exnormal}.

The marginal strategy  tends to be less sensitive to   $\lambda$ than 
the profiled strategy, suggesting relevance  in a \lq\lq conservative" investment   context
with respect to loss function misspecification.  Even with quite widely varying $\lambda$, 
resulting marginal loss portfolios will be more stable, and far less susceptible to substantial changes
and potential deterioration in terms of cumulative returns, than profiled loss portfolios.

\section{Additional Comments \label{sec:conc} } 

The selected illustrations in our application to  financial time series highlight key 
features and benefits of our Bayesian emulation approach to computing optimal decisions,  
as well as our development of new multi-step portfolio decision analysis.   Versions of the
Laplace loss functions generate multi-step portfolios that consistently outperform traditional 
myopic approaches, both with and without transaction costs; they  define psychologically and 
practically attractive framework for investors concerned about portfolio stability over multiple periods
with defined targets.   Examples show the opportunity-- through appropriate selection of 
loss function parameters--  for resulting portfolios to cushion the impacts of 
economically challenging times for the market, and enhance recovery afterwards, as 
highlighted in the examples using FX, commodity and market index data over 2009-2012. 
In addition to showcasing the application of the  concept of   \lq\lq Bayesian emulation for
decisions", our interest in multi-step portfolios also highlights the central question of 
optimization for one-step decisions in the multi-step view; while specific numerical 
methods using dynamic programming might be tuned and customized to a specific loss function
in this context, the Bayesian emulation approach opens up new approaches and suggests
new avenues for development.  As we begin in our discussion of {\em marginal} versus {\em profiled} 
loss functions,  there is now opportunity to drive some part of the research agenda from
synthetic statistical models as a starting point, exploring and evaluating the {\em implied}
loss functions. One specific, current direction linked to this is to define classes of 
non-normal, state-space models with skewed  innovation/error distributions that induce
asymmetric loss functions; a key idea here is to use discrete/mean-scale mixtures of normals for the 
innovation/error distributions,  so maintaining the ability to use MCMC coupled with FFBS/EM
methods for mode-finding while generating a very rich class of resulting loss functions. 
One key and desirable feature of the latter, in particular, is to represent high penalties on 
portfolio short-fall relative to moderate or expected gains.  This direction, and 
others opened-up by the \lq\lq Bayesian emulation for decisions" approach,  offers potential for
impact on research frontiers in statistics and decision theory as well as  application
in financial portfolio development and other areas.


\appendix

\section{Appendix:   Mode Searching for Marginal Laplace \label{sec:appm} }

\subsection{Gibbs Sampler and Maximization of Mixture of Normal Densities   }

To construct the approximate density in \eqn{mix}, 
we need to sample from the (joint) posterior of the model in \eqn{dlm2}. 
The Gibbs sampler for this model has components related to those of Bayesian lasso 
regression~\citep{park2008bayesian},  but now in the extended context of dynamic models 
using FFBS methods.  The MCMC proceeds over iterations $i=1{:}I$ as follows: 
\begin{itemize}
\item Sample each   $\tau_{jt}^{(i)}$ from its generalized inverse Gaussian (GIG)\footnote{The density of $GIG(x|a,b,c)$ is  $ p(x|a,b,c) \propto x^{a-1} \exp \{ (bx + c/x) /2 \}.$}, complete 
conditional posterior distribution, 
$\tau_{jt}^{(i)}|w_{1{:}h}^{(i-1)} \sim  GIG(1/2, 1/\lambda_t^2,(w_{jt}^{(i-1)}-w_{j(t-1)}^{(i-1)})^2),$
independently across $j=1{:}k$ and $t=1{:}h.$ 
\item Sample $w_{1{:}h}^{(i)}|\tau_{1{:}h}^{(i)}$ using FFBS.
\item For later use, record the  means and variances of the marginal normal 
posterior $p(w_1|\tau_{1{:}h}^{(i)}) = N(w_1 | m_1^{(i)},C_1^{(i)})$ generated by the above FFBS analysis. 
\end{itemize}
With the samples $\{ w_{1{:}h}^{(i)}, \tau_{1{:}h}^{(i)} \}$ and the by-products $\{ m_1^{(i)},C_1^{(i)} \}$, 
the Monte Carlo approximation to $p(w_1)$ is  
$$
 \hat{p}(w_1) = I^{-1} \sum_{i=1:I} N(w_1|m^{(i)}_1,C_1^{(i)}).
$$
The next step is mode-finding in this mixture of normals. Modes satisfy
\begin{equation} \label{eq:foc}
w_1 = A_i \sum_{i=1:I} N(w_1|m_i,C_i) C_i^{-} m_i \quad\textrm{where}\quad
A_i^{-1} = \sum_{i=1:I} N(w_1|m_i,C_i) C_i^{-}
\end{equation} 
with $m_i=m_1^{(i)}$ and $C_i = C_1^{(i)}.$ 
We iterate this fixed-point equation to compute approximate modes, with the strategy for multiple \lq\lq global" 
 starting values as noted in the main paper.  Normal mixtures can exhibit multiple modes, and our
starting values-- using means $m_i$ prioritized by the resulting values of $\hat{p}(m_i)$--
explicitly address this by defining a set of \lq\lq spanning" mode searches.

\subsection{Sum-to-One Constraint}

Note that, when the sum-to-one constraint is imposed on the original model by setting $W_t = \mathrm{diag}(\tau_t) - \tau_t \tau_t' / 1_k'\tau_t$, 
then the full conditional of the $\tau_{jt}$ is no longer a product of univariate GIG distributions.
To sample each $\tau_t$, we therefore use a novel, independence chain Metropolis-Hastings algorithm. 
Here the product of the initial GIG distributions in the unconstrained model is used as the obvious 
proposal distribution.  Acceptance probabilities involve singular normal densities based on the
generalized inverses  
$W_t^-=\mathrm{diag}(\tau_t) - \tau_t \tau_t' / (1_k'\tau_t+c)$ where $c=10^{-9}$. 
The acceptance probability of each proposed sample $\tau_{jt}^{new}$ conditional on its previous 
value $\tau_{jt}^{old}$ and all other parameters is 
$(c + 1_k' \tau_{jt}^{new})^{1/2}/(c + 1_k' \tau_{jt}^{old})^{1/2}$
where $\tau_{jt}^{new} = (\tau_{1t},\cdots , \tau_{jt}^{new} , \cdots ,\tau_{kt} )'$ and 
$\tau_{jt}^{old} = (\tau_{1t},\cdots , \tau_{jt}^{old} , \cdots ,\tau_{kt} )'$. 
We observe in our empirical studies and the application of the paper, in particular,  
that the acceptance probability is generally very high-- typically around 98\%. 

We note also that, due to  sum-to-one constraint, the conditional, $k\times k$ variance matrices 
$C_i$ are rank-deficient, being of rank $k-1$.  The generalized inverse
$C_i^{-}$ in~\eqn{foc} are based on singular value decompositions in this iterative 
numerical solver for the modes of the normal mixture.

\section{Appendix: Dynamic Dependence Network Models \label{sec:ddn}}

For time series analysis and forecasting, we adapt the framework of dynamic dependence
network models (DDNMs) introduced in~\cite{ZhaoXieWest2015}. This model framework builds
on prior work in multivariate dynamic modelling and innovates in bringing formal and adaptive Bayesian
model uncertainty analysis to parsimonious, dynamic graphical structures of real practical relevance to
financial (and other) forecasting contexts. Specific classes of DDNMs represent both lagged and
cross-sectional dependencies in multivariate, time-varying autoregressive structures, with an
ability to adapt over time to dynamics in cross-series relationships that advances the ability to
characterize changing patterns of feed-forward relationships and of multivariate volatility, and to
potentially improve forecasts as a result.

Denote the $k\times 1$ vector of assets by $y_t$; in our application, $y_t$ is the vector of 
log prices of the financial assets.  The DDNM extension of a TV-VAR(2) model represents $y_t$ via 
$$
(I_k-\Gamma_t) y_t \sim N(\Phi_tx_t, D_t )
$$
where  $x_t = (1,y_{t-1}',y_{t-2}') '$,   $\Phi_t$ is a $k\times (1+2k)$ matrix of time-varying intercept 
auto-regressive coefficients, $\Gamma_t$ is a time-varying, lower triangular matrix with diagonal zeros, and 
$D_t = \mathrm{diag}(v_{1t},\dots ,v_{kt})$ with time-varying univariate volatilities on the diagonal. 
The model can be written element-wise as 
\begin{equation} \label{eq:mdm}
 y_{jt}  = x_t' \phi_{jt} + y_{pa(j),t}'\gamma_{jt} + N(0,v_{jt}), 
\end{equation}
where $\phi_{jt}'$ is the $j$-th row of  $\Phi_t$, 
$pa(j) \subseteq  \{1{:}(j-1)\}$ is the {\em parental set} of series $j$ defined as the indices of 
$j$-th row of $\Gamma_t$ with non-zero elements, and 
$y_{pa(j),t}$ and $\gamma_{jt}$ are the corresponding subvectors with $|pa(j)|$ elements of 
$y_t$ and $j$-th row of $\Gamma_t$. 
The state parameters $(\phi_{jt},\gamma_{jt})$ are assumed to follow 
normal random walks with a discount factor method applied to define the 
state evolution variance matrices as is standard in univariate DLMs~\citep[][chap.~6]{West1997}. 
The observational variance $v_{jt}$ is modeled as a gamma-beta  stochastic volatility 
process over time, again based on standard DLM methodology~\citep[][sect.~10.8]{West1997}.
Sparsity of the parental sets $pa(j)$ defines patterns of zeros below the diagonal in 
$\Gamma_t$. This in turn defines the sparsity structure of the implied residual 
precision matrix; by inversion, the conditional precision matrix of $y_t$ given
the past values and all dynamic parameters is $(I_k-\Gamma_t')D_t^{-1}(I_k-\Gamma_t)$ 
which has the form of  sparse Cholesky decomposition when $\Gamma_t$ is sparse. If the
level of sparsity in parental sets  is high, then this precision matrix will also have zeros in some of the 
off-diagonal elements, representing conditional independencies in the innovations; since 
elements of $\Gamma_t$ and $D_t$ are time-varying, these conditional independencies 
represent an underlying dynamic graphical model for the innovations. 

Given the parental sets $pa(j)$, the sequential, forward filtering analysis of the multivariate 
DDNM partitions into a parallel set of  $k$ univariate models with standard, analytic 
computation of on-line prior-to-posterior updating and one-step ahead forecasting.  
For forecast distributions more than one-step ahead, simulation methods are used as
the unknown future observations $y_{t+i}$ ($1\le i\le h$) are required as conditional
predictors. Direct simulation from the exact predictive distributions is easily implemented
recursively, as detailed in~\cite{ZhaoXieWest2015}

As in much of our past work in practical financial time series forecasting, we apply the DDNM to 
log prices  in the vector $y_t$, and returns $r_t$ are then inferred. For univariate series $j$ with
price $p_{jt}$ at time $t,$   the return is $r_{jt} = (p_{jt}-p_{j(t-1)})/p_{j(t-1)}$ with
$p_{jt}=\exp(y_{jt})$.   Similar relationships define the $k$-step ahead returns at any time. 
Our portfolio analyses require predictive mean vectors and variance matrices of returns, which
can be directly computed by transformation of the predictive samples of log prices.

The DDNM requires specification of the parental sets $pa(j)$. We choose these 
based on exploratory analysis of preliminary data over first 500 days. Filtering and forecasting with the
defined DDNM then run from day 501, redefined as $t=0$ in the formal sequential analysis. 
This exploratory analysis runs {\em full} models over the first 500 days, i.e., DDNMs using 
$pa(j) = 1{:}(j-1)$ for each $j=1{:}k.$  Then,  we simply compute the 
Cholesky decomposition of the posterior mean of $\Gamma_{500}$ and 
threshold its off-diagonal elements using a threshold of $d=0.2;$ those elements exceeding the
threshold in row $j$ define the parental set $pa(j)$ (with, of course,  $pa(1) = \emptyset)$ that
we adopt for the forward filtering and forecasting analysis from then on.   
The choice of the threshold is naturally important here; a higher threshold yields sparser 
parental sets and hence sparser $\Gamma_t$ matrices.   Exploratory analysis on the first 
500 days is used to explore and evaluate this, and guide the choice informally.  With a very low
threshold, forecasts of returns in this training period tends to have very narrow credible intervals 
but show substantial biases, especially in multi-step ahead prediction. 
Higher thresholds-- consistent with increased sparsity-- lead to wider credible intervals but 
less adaptive models.  On a purely exploratory basis, we chose $d=0.2$ as a \lq\lq sweet-spot"
balancing forecast mean accuracy and uncertainty.   This-- ad-hoc but practically 
rationale-- exploratory analysis defined a relevant, specific DDNM for use here. 
The resulting parental sets are displayed in Table~\ref{tab:paj}. 

Our results are based on   the resulting DDNM with additional 
parameters as follows: for the normal DLM state evolutions, we use discount factor 
of 0.98 for each series, and 0.97 for residual stochastic volatilities. Direct simulation of
multi-step ahead predictive distributions used a Monte Carlo sample size of  50{,}000.

\begin{table}[!htbp]
\begin{center}
 \begin{tabular}{lc} 
Parent $j$ & $ pa(j)$ \\  
 \phantom{.} \\
OIL & $\emptyset$ \\
GBP & $\emptyset$ \\
EUR & $\emptyset$ \\
NOK & EUR  \\
ZAR & GBP NOK  \\
CAD & $\emptyset$ \\
AUD & NOK CAD \\ 
NZD & AUD  \\
JPY & GBP EUR CAD AUD  \\
CHF & $\emptyset$ \\
GLD & GBP ZAR CAD CHF \\ 
S\&P& GBP EUR NOK CAD AUD NZD \\ 
NSD & AUD JPY CHF S\&P \\
 \end{tabular}
 \caption{\small Parental sets used in prediction.} \label{tab:paj}
\end{center}
\end{table}%

\end{document}